\documentclass{article}

\usepackage[main, final]{neurips_2025}

\usepackage[utf8]{inputenc} 
\usepackage[T1]{fontenc}    
\usepackage{hyperref}       
\usepackage{url}            
\usepackage{booktabs}       
\usepackage{amsfonts}       
\usepackage{nicefrac}       
\usepackage{microtype}      
\usepackage{xcolor}         
\usepackage{tabularx}
\usepackage{graphicx}
\usepackage{listings}
\usepackage{amsmath}
\usepackage{subcaption}
\usepackage{multirow}

\lstdefinestyle{excel}
{
    basicstyle=\footnotesize\ttfamily,
    string=[d]{'},
    stringstyle={\color{gray}},
    escapeinside={(*}{*)},
    showspaces=false,
    showstringspaces=false,
}

\newcommand{\circled}[1]{\scalebox{1.1}{\raisebox{.4pt}{\textcircled{\raisebox{-.4pt} {\scalebox{.8}{#1}}}}}}

\newcommand{\cx}[1]{\mathbf{x}_{#1}}
\newcommand{\ptheta}{p_\theta}

\DeclareMathOperator*{\argmin}{arg\,min}
\DeclareMathOperator{\ce}{ce}

\newtheorem{example}{Example}

\title{Diffusion is a code repair operator and generator}

\author{%
  Mukul Singh \\
  Microsoft\\
  Redmond, WA\\
  \texttt{singhmukul@microsoft.com} \\
  \And
  Gust Verbruggen \\
  Microsoft \\
  Belgium \\
  \texttt{gverbruggen@microsoft.com} \\
   \AND
  Vu Le \\
  Microsoft \\
  Redmond, WA \\
  \texttt{levu@microsoft.com} \\
  \And
  Sumit Gulwani \\
  Microsoft \\
  Redmond, WA \\
  \texttt{sumitg@microsoft.com}
}

\begin{document}

\maketitle

\begin{abstract}
Code diffusion models generate code by iteratively removing noise from the latent representation of a code snippet.
During later steps of the diffusion process, when the code snippet has almost converged, differences between discrete representations of these snippets look like last-mile repairs applied to broken or incomplete code.
We evaluate the extent to which this resemblance can be exploited to leverage pre-trained code diffusion models for the problem of last-mile repair by considering two applications with significant potential.
First, we can leverage the diffusion model for last-mile repair by adding noise to a broken code snippet and resuming the diffusion process.
Second, we can leverage the diffusion model to generate arbitrary amount of training data for last-mile repair tasks (that are computationally more efficient) by sampling an intermediate program (input) and the final program (output) from the diffusion process.
We perform experiments on 3 domains (Python, Excel and PowerShell) to evaluate applications, as well as analyze properties. \footnote{The code and associated datasets can be found at \texttt{redacted}}
\end{abstract}

\section{Introduction}

Diffusion models have emerged as a powerful paradigm in generative modeling, particularly for tasks that involve complex data structures \citep{diffusion}.
Instead of generating a sample from a distribution in one go (like a GAN or VAE) or auto-regressively (like a GPT) they learn to iteratively reverse diffusion steps that add (typically Gaussian) noise to the data.
Initially popularized in the domain of image generation, diffusion models have since been adapted for modalities like video generation \citep{videodiffusion,videodiffusionsurvey}---which requires a temporal component---and text or code generation \citep{diffusionlm,codefusion}---which requires diffusion over discrete tokens.

One approach of applying diffusion to discrete domains, like text or code, involves embedding the input, performing diffusion in the embedded representation, and projecting the denoised embeddings back to discrete tokens.
To train this model end-to-end, the loss incorporates a component over the discrete tokens, meaning that representation from each step of the reverse diffusion process can be converted back to the discrete space \citep{genie}.
During initial generations, decoding the latent representation does not resemble anything and tokens frequently change, but in later generations, these decoded representations become readable and it takes multiple steps to change one token.

As an example, consider the following generations from pre-trained CodeFusion \citep{codefusion} models---without natural language conditioning---trained on Excel
\begin{center}
    \begin{tabularx}{.9\textwidth}{rl}
    $(t_{75\%})$& {\ttfamily\small =IF(COUNTIF(A:A, `>10')=0, `No values', AVERAGE(A:A))} \\ 
    $(t_{90\%} - t_{100\%})$& {\ttfamily\small =IF(COUNTIF(A:A, `>10')=0, `No values', AVERAGE\textcolor{red}{IF}(A:A\textcolor{red}{, `>10'}))}\\
\end{tabularx}
\end{center}
and Python
\begin{center}
    \begin{tabularx}{.9\textwidth}{rl}
    $(t_{75\%} - t_{90\%})$& {\ttfamily\small words = read(`myfile').split()} \\ 
   $(t_{100\%})$& {\ttfamily\small words = open(`myfile')\textcolor{red}{.read()}.split()}
\end{tabularx}
\end{center}
with changed tokens highlighted in \textcolor{red}{red}.
It appears as if the diffusion model can look at the whole (discrete) program, determine what is missing to make it functional, and apply those fixes.
This is exactly the premise of last-mile repair, in which the goal is to repair broken code in such a way that the solution differs minimally from the broken code \citep{lamirage}.
A major challenge in training (last-mile) repair systems is the long-tail problem in obtaining training data \citep{repairfinetuning} and out-of-distribution generalization when introducing synthetic errors \citep{flame}.

In this paper, we address those challenges by evaluating the extent to which discrete token changes during reverse diffusion steps are representative of last-mile repair fixes.
This evaluation is broken down in two main applications: whether we can use the diffusion model to \textbf{directly repair code}, and whether we can use the diffusion model to \textbf{generate training data} for fine-tuning repair models.

We support our claims with experiments on three programming languages: Python, PowerShell and Excel. We find that diffusion models are capable of last-mile repair, with the models being able to repair 56.4--68.2\% of Python and Excel snippets across different noise levels. 
We also find that the diffusion-generated synthetic data has higher diversity and complexity compared to existing  data generators and GPT-4o, which is reflected in higher performance observed (+2.5 -- 3.5\%) when fine-tuning different models (\texttt{\small codet5-small}, \texttt{\small phi-35-mini} and \texttt{\small mistral-7b}) on the synthetic data.

\begin{figure*}
    \centering
    \includegraphics[width=\textwidth]{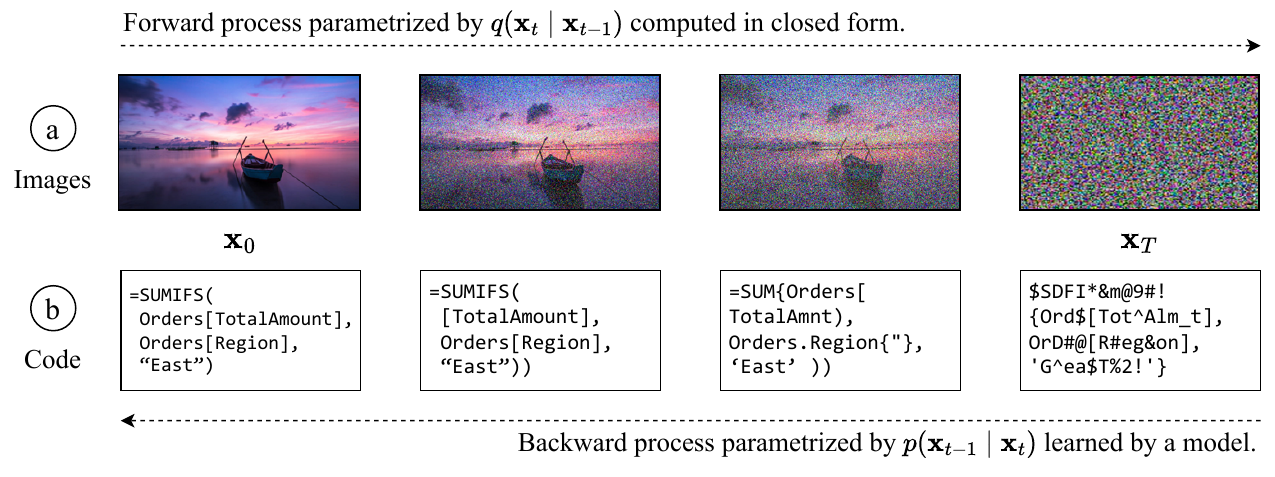}
    \caption{Example of diffusion for \circled{a} images and \circled{b} code. Pure $\cx{T}$ is iteratively denoised into a sample $\cx{0}$ from the target distribution by a model trained on data from the forward process.}
    \label{fig:background}
\end{figure*}

\section{Background}
\label{sec: background}

\subsection{Diffusion models}

A diffusion model is a latent variable model that constructs a Markov chain $\cx{0}, \cx{1} \cdots \cx{T}$ and simulates data $\cx{0} \sim p_\text{data} $ by learning to reverse this Markov chain \citep{diffusion}.
The sequence of continuous latent variables $\cx{1:T}$ is constructed by incrementally adding (typically Gaussian) noise to data $\cx{0}$ until, at diffusion step $T$, samples $\cx{T}$ are approximately Gaussian.
Each transition $\cx{t-1} \rightarrow \cx{t}$ is parametrized by $q(\cx{t} \mid \cx{t-1}) = \mathcal{N} (\cx{t} ; \sqrt{1-\beta_t} \cx{t-1}, \beta_t \mathbf{I})$ where the hyper-parameter $\beta_t$ is the amount of noise added at diffusion step $t$. 
The diffusion model generates samples by reversing this chain: it iteratively denoises the sequence of latent variables $\cx{T:0}$ to approximate a sample from the target distribution. Each denoising transition $\cx{t} \rightarrow \cx{t-1}$ is parametrized by the model that predicts $\ptheta(\cx{t-1}\mid \cx{t}) = \mathcal{N}(\cx{t-1}; \mu_\theta(\cx{t}, t), \Sigma_\theta(\cx{t}, t))$. 
In practice, instead of constructing the whole chain, we can immediately obtain $\cx{t}$ from $\cx{0}$ as $\cx{t} = \sqrt{\bar{\alpha}_t}\cx{0} + \sqrt{1 - \bar{\alpha}_t}\epsilon$ with $\bar{\alpha}_t = \prod_{i=1}^t 1 - \beta_t$ and $\epsilon \sim \mathcal{N}(0, \mathbf{I})$.
The model $f_\theta(\cx{t}, t)$ is parametrized to predict $\cx{0}$ with an empirically validated loss function $\mathcal{L}_{\mathrm{simple}} = \mathbb{E}_{\cx{0}, \epsilon_t, t} \Vert f_\theta(\cx{t}, t) - \cx{0} \Vert^2$ \citep{diffusion,diffusionlm}.
At inference time, we compute $\cx{t-1} = \sqrt{\bar{\alpha}_t}f_\theta(\cx{t}, t) + \sqrt{1 - \bar{\alpha}_t}\epsilon$ to iteratively denoise $\cx{t}$.

\begin{example}
    Figure~\ref{fig:background} shows the generations of a latent diffusion model. It can be seen how the model iteratively denoises to the concrete representation from the output space.
\end{example}

\subsection{Diffusion models for code}

Code generation is a discrete generation task, where the expected output is a snippet $\mathbf{c} = [c_1, \ldots, c_k]$ of $k$ tokens.
CodeFusion \citep{codefusion} draws inspiration from text diffusion \citep{diffusionlm} where each token $c_i$ is embedded $\textsc{e}(c_i) \in \mathbb{R}^d$ to convert $\mathbf{c}$ into a continuous representation $\textsc{e}(\mathbf{c}) \in \mathbb{R}^{kd}$ to which a regular diffusion process can be applied.
In the reverse process, a trainable rounding step $p_\theta(c_i \mid x_{\leq i})$ computes a distribution over possible tokens for each position $i$ given all previous (denoise) tokens $x_{\leq i}$.
Note that the decoder is trained to always generate a constant number of $n > k$ tokens, one of which is an end-of-sequence token and $n - k - 1$ padding tokens.
Like CodeFusion, we set $n = 128$.

\begin{example}
    Figure~\ref{fig:background} shows the generations of a latent code diffusion model.
    The intermediate representations, when visualized in the discrete token space, show how the model iteratively denoises to a syntactically valid Excel formula. Furthermore, we can see how the generation at $t_{75\%}$ has the table name missing in the structured reference which the model fixes through refinement.
    \begin{center}
        \begin{tabularx}{.9\textwidth}{rl}
            $(t_{75\%})$& {\ttfamily\small =SUMIFS([TotalAmount], Orders[Region], "East"))} \\ 
            $(t_{90\%} - t_{100\%})$& {\ttfamily\small =SUMIFS(\textcolor{red}{Orders}[TotalAmount], Orders[Region], "East"))}\\
        \end{tabularx}
    \end{center}
\end{example}

\begin{figure}[htb]
    \centering
    \begin{subfigure}[t]{0.4\textwidth}
        \centering
        \includegraphics[width=\textwidth]{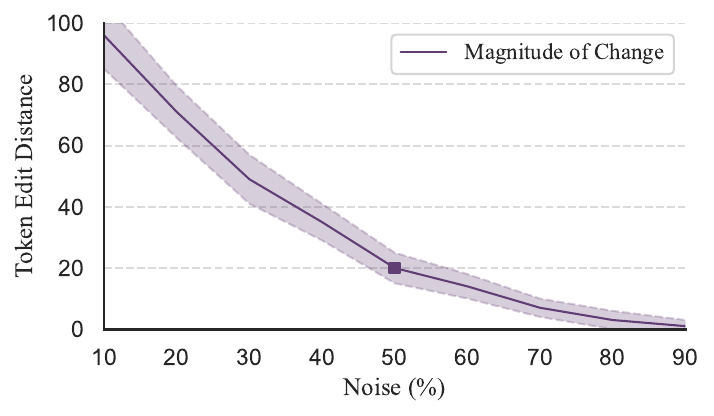}
        \caption{Percentage of tokens changed in each iteration.}
        \label{fig:mag-of-change}
    \end{subfigure}
    \hfill
    \begin{subfigure}[t]{0.4\textwidth}
        \centering
        \includegraphics[width=\textwidth]{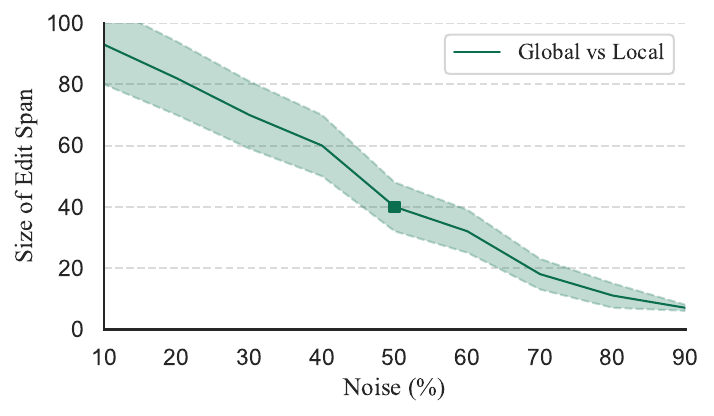}
        \caption{Length of spans of tokens that are changed at once.}
        \label{fig:global-vs-local}
    \end{subfigure}
    \caption{Trends in code refinement over diffusion time steps.}
    \label{fig:diffusion-pattern-analysis}
\end{figure}

More generally, Figure~\ref{fig:diffusion-pattern-analysis} shows trends in discrete code refinement over subsequent diffusion time-steps as (a) the number of tokens being changed and (b) the lengths of spans of tokens that are changed at once.
As expected, significantly fewer tokens are changed further down the diffusion process.
These trends of fewer and localized edits near the end of the diffusion process motivate the application of the diffusion process for last-mile repair.

\begin{figure*}
    \centering
    \includegraphics[width=.9\textwidth]{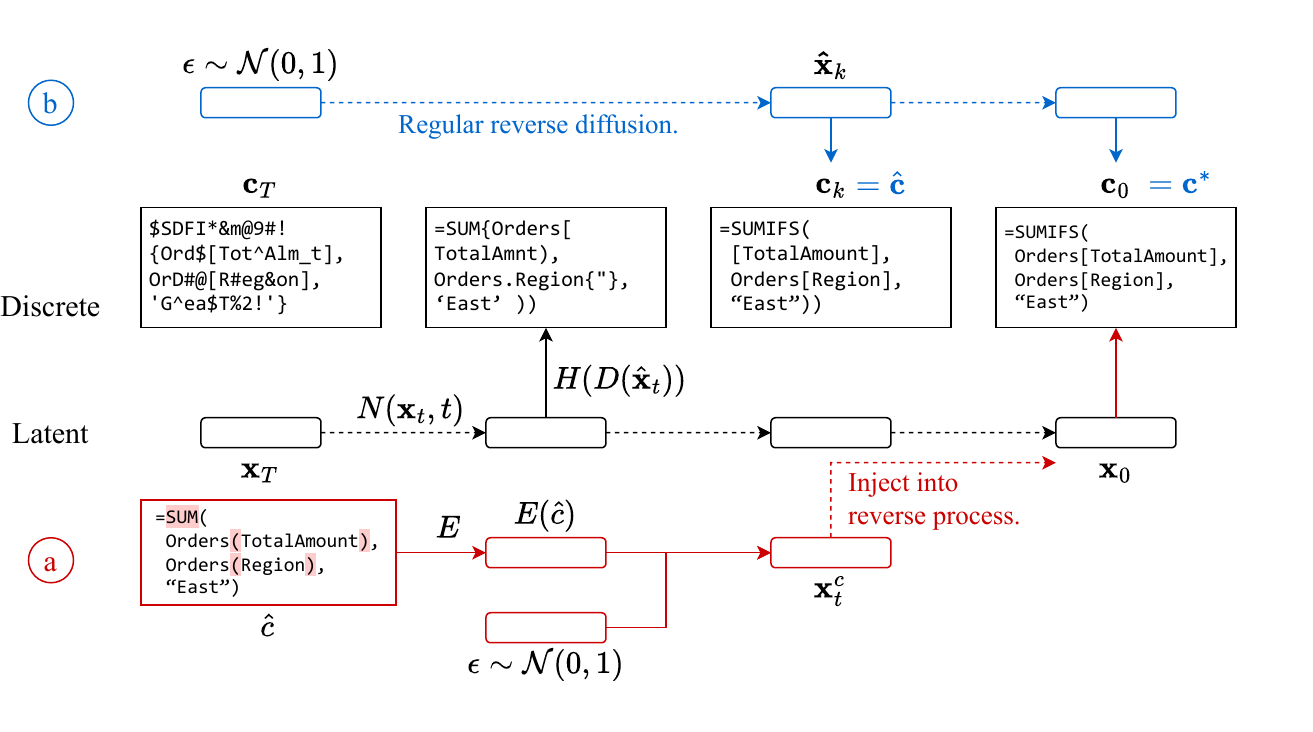}
    \caption{Using a pre-trained diffusion process (in black) to \circled{a} repair broken code and \circled{b} generate (broken, fixed) code pairs for training specialized approaches. \circled{a} The broken code is embedded, noise is added for a timestep $t$, and the reverse process is resumed as usual, letting the reverse process fix the code. \circled{b} The diffusion process produces intermediate (broken) code snippets $\hat{\mathbf{c}}$ that can be paired with the final code $\mathbf{c}^*$ to form a training example.}
    \label{fig:method}
\end{figure*}

\section{Diffusion for repair}

Let $\hat{c}$ be a buggy code snippet that is not accepted by the compiler.
The goal of last-mile repair is to find a code snippet $c^* = \argmin_{c} d(c, \hat{c})$ such that $c^*$ is accepted by the compiler and performs a task intended by the user, with $d$ the edit distance between two code snippets.
Like previous work on last-mile repair, we only consider syntactic errors \citep{lamirage,ring}.
In the following three sections, we respectively reiterate the components and training process of \textsc{CodeFusion}, describe how to apply it to problem of last-mile repair, and describe how to generate pairs $(\hat{c}, c^*)$ that can be used to train specialized systems.

\subsection{Training the diffusion model}

The pre-trained components of \textsc{CodeFusion} generate code from pure Gaussian noise.
Because there is no natural language, we can remove the encoder.
A denoiser $N$ removes the noise from $\cx{t}$ at timestep $t$ to obtain the denoised embeddings $\hat{x}_0 = N(\cx{t}, t)$.
A decoder $D$ performs full self-attention over $\hat{x}_0$ to compute a decoded representation $D(\hat{x}_0)$.
This allows each denoised token to be generated with information about other tokens, and improved the likelihood of generating syntactically correct code \citep{codefusion}.
Finally, the classification head $H$ computes $p(y \mid d_i)$ for each $d_i \in D(\hat{x}_0)$ to project decoded embeddings back to discrete tokens.

To train these components on a code snippet $\mathbf{c}$, an embedding layer $E$ first obtains the continuous representation $\cx{0} = E(\mathbf{c})$.
We sample $t \in [1, \ldots, T]$ and $\epsilon_t \sim \mathcal{N}(0, 1)$ and compute $\cx{t}$ from $\cx{0}$.
The model is trained on
\begin{equation*}
    \mathcal{L} = \underbrace{\Vert N(\cx{t}, t) - \cx{0} \Vert}_{1} + \underbrace{\Vert D(\hat{x}_0) - E(\mathbf{c}) \Vert}_{2} - \underbrace{\ce(\mathbf{c}, H(D(\hat{x}_0)))}_{3}
\end{equation*}
and consists of three parts that
\begin{enumerate}
    \item minimize the error between the predicted noise  $\hat{\epsilon}_t$ and the actual noise $\epsilon_t$ to train $N$,
    \item minimize the error between the decoded embeddings $D(\hat{\mathbf{x}}_0)$ and embedded code $E(\mathbf{c})$ to train $D$ and $L$, and
    \item apply cross-entropy loss with respect to the ground truth code snippet $\mathbf{c}$ to train $H$.
\end{enumerate}
This loss is taken from \textsc{CodeFusion} \citep{codefusion} and is an adaptation of the loss function used by \textsc{genie} \citep{genie}.

\subsection{Diffusion steps as repair \emph{operators}} \label{sec:repair}

We exploit the Markov property of the reverse diffusion process to \emph{inject} an embedded version of the noisy snippet into the reverse process.
In other words, we can pick some $t$, generate $\epsilon \sim \mathcal{N}(0, 1)$ and compute $\cx{t}^{\hat{\mathbf{c}}} = \sqrt{\bar{\alpha}}E(\hat{\mathbf{c}}) + \sqrt{1 - \bar{\alpha}_t}\epsilon$ where $E$ is the embedding layer (that CodeFusion discards after training).
The diffusion process then denoises $\cx{t}^{\hat{\mathbf{c}}} \rightarrow \cx{0}^{\hat{\mathbf{c}}}$ and we return $\ H(D(N(\cx{0}^{\hat{\mathbf{c}}}, 0)))$.

Let $\mathbf{X}_{t}^{\hat{\mathbf{c}}}[E]$ be the space of embedded representations $\cx{t}^{\hat{\mathbf{c}}}$ obtained from $\hat{\mathbf{c}}$ for all $\epsilon \sim \mathcal{N}(0, 1)$ at step $t$ (parametrized by $E$).
Let $\mathbf{X}_{t}^{\mathbf{c}^*}[N, D, H]$ be the space of embedded representations encountered at step $t$ in reverse diffusion processes starting from $\epsilon \sim \mathcal{N}(0, 1)$ that end up in $\mathbf{c}^*$ (parametrized by $N$, $D$ and $H$).
Our intuition is that there exists some $t$ for which these spaces have a significant overlap, and there are thus many values of $\epsilon$ that project $\hat{\mathbf{c}}$ into a trajectory to $\mathbf{c}^*$.
If $t$ is too large, the probability of ending up there is small (too much noise).
If $t$ is too small, it will never end up there (not enough noise).

\subsection{Diffusion models as repair \emph{generators}}
\label{sec:synthetic}

We exploit the seemingly discrete  nature of later diffusion steps to generate synthetic repair data.
Starting the reverse process from $\cx{T} \sim \mathcal{N}(0, 1)$ we build the chain $\hat{\mathbf{x}}_T \rightarrow \hat{\mathbf{x}}_0$ and decode each snippet into $\mathbf{c}_T \rightarrow \mathbf{c}_0$.
We can then select any $(\mathbf{c}_t, \mathbf{c}_0)$ as a training pair if $\mathbf{c}_t \neq \mathbf{c}_0$.

In previous work, mistakes are introduced in the discrete token space, by implementing specialized functions that imitate human errors \citep{aprselfsupervised,flame,datavinci, aligned-code-gen} and optionally training a neural network to imitate those \citep{bifi}.
We show that the space of discrete representations encountered during the reverse diffusion process shares enough similarities to the discrete last-mile repair errors.

\section{Experiments}

We evaluate both how the diffusion process acts as a repair operator, how the generated data can be used for supervised repair training, providing insights on how diffusion generates (and repairs) code.

\subsection{Experimental Setup}

\paragraph{Benchmarks}
We evaluate our approach on three different benchmarks that span different types of code (formulas, code, commands).
\begin{enumerate}
    \item \textbf{Excel} \citep{lamirage} is a benchmark of 200 broken formulas collected from a public Excel help forum\footnote{\url{www.mrexcel.com}}.
    \item \textbf{PowerShell} \citep{ring} is a repair benchmark for 208 PowerShell commands collected from StackExchange\footnote{\url{www.stackexchange.com}} by comparing commends in the question with those in accepted answers.
    \item \textbf{Python} \citep{bifi} is a code repair benchmark collected from GitHub. We evaluate on a random sample of 200 syntactically invalid Python code snippets. These do not have a ground truth repair, hence,
    we employ the same evaluation metric described in the BIFI
    paper using (1) syntactic validity and (2) token edit distance $< 5$.
\end{enumerate}

\paragraph{Pre-training data} Collecting snippets of code for unsupervised approaches is significantly easier than finding data for repair.
\begin{enumerate}
    \item For \textbf{Python}, we use a collection of 130K code snippets for simple tasks with an average token length of 79.4 tokens from existing benchmarks \citep{bifi}.
    \item For \textbf{Excel}, we use a corpus of 1.8 million workbooks \citep{cornet, payan2023instructexcelbenchmarknaturallanguage}, and sample 200K workbooks and collect all formulas present in them to generate 108K unique formulas with an average length of 35.8 tokens.
    \item For \textbf{PowerShell}, we collect PowerShell commands from public websites. The corpus has 110K samples with an average length of 24.9 characters.
\end{enumerate}

\paragraph{Metrics}

When available, we use execution match---comparing the output of executing the repaired code with an expected output---which allows for semantically different but functionally equivalent code snippets.
To further analyze the syntactic closeness of the repairs to the original code, we also report sketch match, which is implemented as the exact string match of code with constants (strings, numbers, cell references) anonymized.

\paragraph{Models} We implement the same architecture and pre-training as CodeFusion \citep{codefusion}.
The \textbf{embedding} ($E$) has a dimension of 512.
The \textbf{denoiser} ($D$) is a transformer encoder \citep{transformer} with 10 transformer blocks.
The \textbf{decoder} ($D$) is a block with 6 transformer decoder layers.
The \textbf{classification head} ($H$) is a single fully connected layer.

\paragraph{Compute} All training was done on a 8 x A100 cluster (Azure \texttt{NDm\_A100\_v4}).

\subsection{Diffusion for code repair}

\begin{table}[tb]
\centering
\caption{Repair results for CodeFusion. We report sketch and execution match for all languages at different noise pooling settings.\\
}
\label{tab:repair}
\small
\begin{tabular}{lcccccc}
\toprule
& \multicolumn{2}{c}{Python}         & \multicolumn{2}{c}{PowerShell}            & \multicolumn{2}{c}{Excel}  \\ \cmidrule(lr){2-3} \cmidrule(lr){4-5} \cmidrule(lr){6-7}
Noise pooling & Sketch & Execute & Sketch & Execute & Sketch & Execute \\ \midrule
any\%    & 65.3 & 68.1 & 14.3 & 21.2 & 62.3 & 63.4 \\
best\%   & 60.4 & 62.0   & 11.0   & 17.4 & 56.2 & 58.9 \\
vote\% & 61.2 & 62.4 & 11.7 & 18.2 & 57.1 & 59.1 \\
\bottomrule
\end{tabular}
\end{table}

\begin{figure}[tbh]
    \centering
    \begin{subfigure}[b]{0.325\textwidth}
        \centering
        \includegraphics[width=\textwidth]{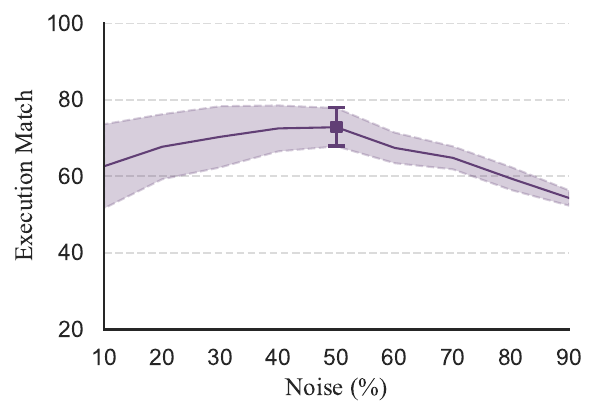}
        \caption{Excel}
        \label{fig:repair-excel}
    \end{subfigure}
    \hfill 
    \begin{subfigure}[b]{0.325\textwidth}
        \centering
        \includegraphics[width=\textwidth]{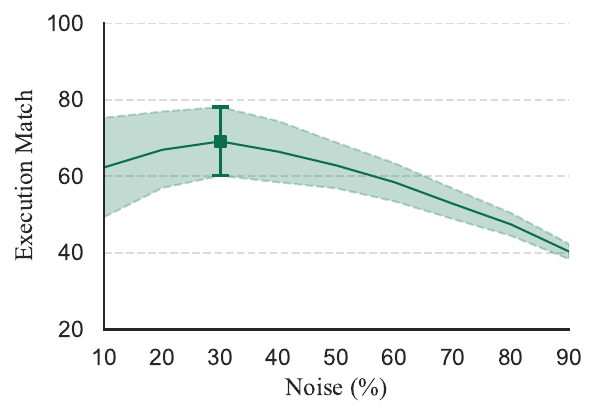}
        \caption{Python}
        \label{fig:repair-python}
    \end{subfigure}
    \hfill 
    \begin{subfigure}[b]{0.325\textwidth}
        \centering
        \includegraphics[width=\textwidth]{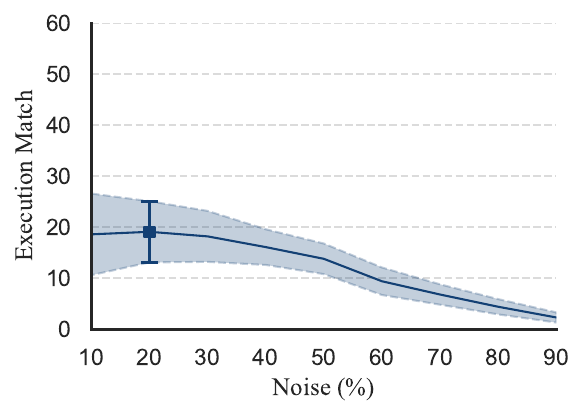}
        \caption{PowerShell}
        \label{fig:repair-powershell}
    \end{subfigure}
    \caption{The evolution of execution match for increasing noise levels added to the noisy snippet ($\hat{\mathbf{c}}$). The optimal noise level is marked. We find that for simpler languages like formulas, injecting later helps while for more complex languages like Python and PowerShell, injecting earlier gives the model more time to repair to the correct code.}
    \label{fig:repair}
\end{figure}

We evaluate a pre-trained diffusion model on last-mile repair.
Table~\ref{tab:repair} contains the execution and sketch match, pooled across noise levels using three strategies, for different diffusion architectures.
In \textbf{any\%}, any noise level was able to correctly repair the code for each sample, indicating the promise of diffusion for repair.
In \textbf{best\%}, we pick the best global noise level for each benchmark set, which are indicated in Figure~\ref{fig:repair}.
In \textbf{vote\%}, we pick the repaired code that was obtained most often across noise levels (using exact string match).
Our findings show that vote-pooling across noise levels is \emph{slightly} more effective than the \emph{free lunch} of an optimal noise level.

Figure~\ref{fig:repair} shows how the execution match evolves in function of the noise level (increments of 10\%) and marks the ``optimal'' noise level (based on average + one standard deviation).
Last-mile repairs are typically small, causing all lower noise levels to work.
For larger noise levels, we see a decline in performance, as the model makes too many changes to the code.

\begin{figure}[tb]
    \centering
    \begin{subfigure}[b]{0.325\textwidth}
        \centering
        \includegraphics[width=\textwidth]{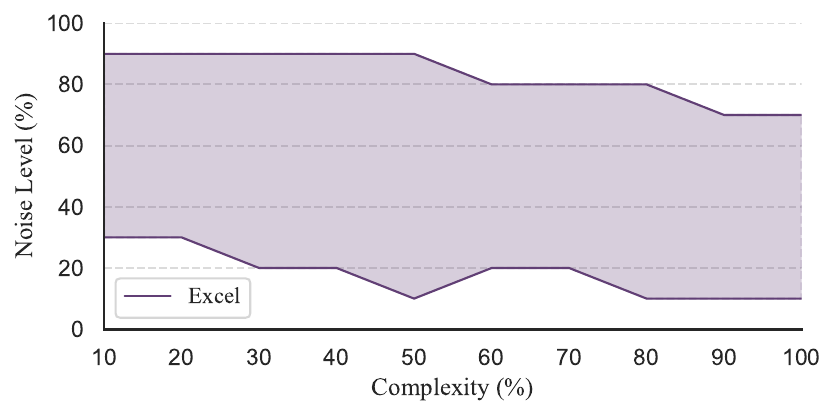}
        \caption{Excel}
        \label{fig:noise-excel}
    \end{subfigure}
    \hfill 
    \begin{subfigure}[b]{0.325\textwidth}
        \centering
        \includegraphics[width=\textwidth]{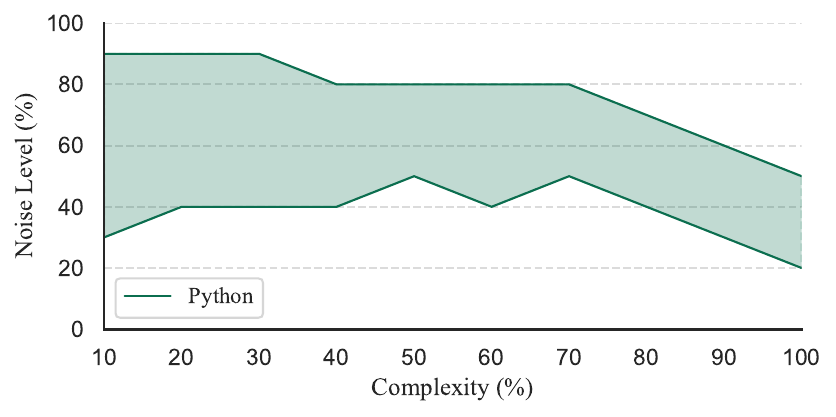}
        \caption{Python}
        \label{fig:noise-python}
    \end{subfigure}
    \hfill 
    \begin{subfigure}[b]{0.325\textwidth}
        \centering
        \includegraphics[width=\textwidth]{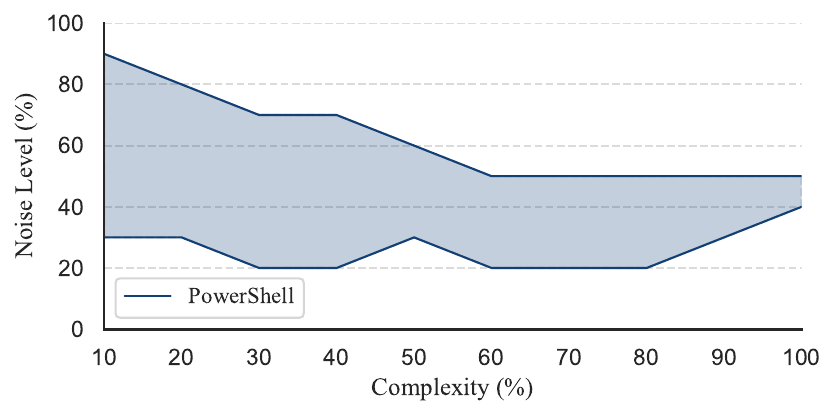}
        \caption{PowerShell}
        \label{fig:noise-powershell}
    \end{subfigure}
    \caption{Noise range for which the correct code snippet is recovered for increasing differences between the broken and fixed code.
    We show the maximum and minimum noise for which a sample was repaired correctly.
    The band width is largest for Excel since it requires simpler and fewer changes while PowerShell has a narrow band towards higher noise as it needs more iterations to repair.}
    \label{fig:repair-complexity}
\end{figure}

Additionally, we examine how error complexity correlates with the noise levels required for repair. Figure~\ref{fig:repair-complexity} shows an area plot of maximum and minimum noise levels where the correct code is generated at least once with increasing complexity, computed as normalized edit distance for the repair tasks.
The results suggest the acceptable noise band varies based on the complexity where an earlier injection is preferred for more complex tasks as these require more iterations to repair.
Furthermore, across languages, we see that Excel has a much wider band as it requires fewer edits while for Python and PowerShell more edits are required for the repair.

To put our results in perspective, Table~\ref{tab:repair-baselines} compares the pass@1 rate of the pre-trained diffusion model with existing approaches.
For LaMirage \citep{lamirage} and BIFI \citep{bifi} we report the numbers from their respective papers.
For the other approaches, we re-implement them, with the Codex \citep{codex} and GPT-4o results based on the \textsc{ring} prompt without compiler feedback.
We note that these very powerful models (GPT-4o), specific repair systems (BIFI and LaMirage) and using additional context (\textsc{ring}) still perform better.
Still, outperforming the Codex model on Python (+8\%) and PowerShell (+11\%) with a small (60M parameter) model that was not specifically trained for repair, is a remarkable result that indicates significant potential of applying diffusion to code repair.

\begin{table}[htb]
\centering
\small
\caption{Comparison performance of CodeFusion with state-of-the-art last-mile repair approaches.\\[1em]}
\label{tab:repair-baselines}
\begin{tabular}{llcrrrrrr}
\toprule
               &&      & \multicolumn{2}{c}{Python} & \multicolumn{2}{c}{Excel} & \multicolumn{2}{c}{PowerShell} \\
\cmidrule{4-5} \cmidrule{6-7} \cmidrule{8-9}
Approach       &Type& Year & Sketch     & Exec.     & Sketch     & Exec.    & Sketch  & Exec.       \\
\midrule
Codex          &Prompt& 2021 & 0.56       & 0.60          & 0.65       & 0.67         & 0.08         & 0.10            \\
RING           &Prompt& 2022 & 0.78       & 0.82          & 0.68       & 0.74         & 0.15         & 0.18            \\
GPT-4o         &Prompt& 2024 & 0.81       & 0.84          & 0.68       & 0.75         & 0.15         & 0.24            \\
GPT-4.1         &Prompt& 2025 & 0.82       & 0.83          & 0.67       & 0.76         & 0.14         & 0.24            \\
GPT-o1         &Prompt& 2025 & 0.81       & 0.82          & 0.65       & 0.74         & 0.12         & 0.21            \\
LaMirage       &Fine-tuned& 2022 & 0.67       & 0.71          & 0.69       & 0.72         & --           & --              \\
BIFI           &Fine-tuned& 2021 & 0.72       & 0.76          & --         & --           & --           & --              \\
CodeFusion     &Pre-trained& -- & 0.65       & 0.68          & 0.62       & 0.63         & 0.14         & 0.21            \\
\bottomrule
\end{tabular}
\end{table}

A major advantage of CodeFusion is its ability to generate diverse outputs, as it is conditioned on noise.
Table~\ref{fig:passk} shows the pass@1, pass@3 and pass@5 rates for diffusion, GPT-4o and \textsc{ring}.
CodeFusion sees the biggest jump in performance ($\pm$ 5\%) across all languages, even performing better than GPT-4o on the (most difficult) PowerShell benchmark.
This reinforces the potential of diffusion for last-mile repair.
Pooling over different noise vectors, execution feedback and larger diffusion models can leverage this potential even further, which we leave for future work.

\begin{table}[htb]
\centering
\small
\caption{Pass@$k$ rates for repair for the best diffusion model adapted for repair (CodeFusion). The performance for CodeFusion increases with $k$, as it generates diverse repairs due to its noise condition.\\}
\label{fig:passk}
\begin{tabular}{lrrrrrrrrr}
\toprule
                       & \multicolumn{3}{c}{Python} & \multicolumn{3}{c}{Excel} & \multicolumn{3}{c}{PowerShell} \\ \cmidrule(lr){2-4} \cmidrule(lr){5-7} \cmidrule(lr){8-10}
Approach               & p@1  & p@3  & p@5  & p@1  & p@3  & p@5  & p@1  & p@3  & p@5 \\ \midrule
CodeFusion             & 68.1 & 70.5 & 72.4 & 63.4 & 65.8 & 68.2 & 21.2 & 23.1 & 26.4 \\
GPT-4o & 81.2 & 81.7 & 82.1 & 75.3 & 75.6 & 75.7 & 23.9 & 24.1 & 24.2 \\
GPT-4.1 & 81.3 & 81.6 & 82.2 & 75.4 & 75.7 & 75.7 & 23.8 & 24.2 & 24.3 \\
GPT-o1 & 81.1 & 81.4 & 82.0 & 75.3 & 75.4 & 75.5 & 23.5 & 24.0 & 24.2 \\
\textsc{ring} & 82.4 & 82.6 & 82.9 & 73.8 & 74.2 & 74.5 & 18.0 & 18.0 & 18.2 \\
\bottomrule
\end{tabular}
\end{table}

\subsection{Diffusion for synthetic data generation}

We evaluate the pre-trained diffusion model (CodeFusion architecture) on generating training data for supervised approaches.
We uniformly sample $t$ and select $(\mathbf{c}_t, \mathbf{c}_0)$ from the diffusion process. 
We then fine-tune several code generation models on this dataset and evaluate their performance on a repair benchmark containing real human errors.
We sample 20K training points.

As baselines, we consider generators from existing work, as well as generating data with a large language model (GPT-4o).
For Python, we use the popular BIFI \citep{bifi} model, which learns to break code based on a set of manually curated repair operators.
For Excel, we use the 17 operators used to fine-tune \texttt{FLAME} \citep{flame} on last-mile formula repair.
The prompt for GPT-4o is a few-shot, chain-of-thought prompt where we instruct the model to break a formula according to mistakes that a human would make.
We use two versions: (1) using error categories from BIFI to mimic human errors and (2) not providing guidance to promote diversity in mistakes.

Table~\ref{tab:synthetic-data} shows the performance of different data generation techniques across various models: CodeT5+ (2B) \citep{codet5+}, Phi-3.5-mini-instruct (3.8B) \citep{phi3} and Mistral-7B-instruct-v0.3 (7B) \citep{mistral}.
Our results show that models trained on diffusion-generated consistently outperform with other specialized approaches, across all models.
Similar to repair, a significant contributor is the diversity in the generated data, which is harder to control for GPT-4.1.

We analyze properties of the generated data distributions for diffusion and GPT-4o in Figure~\ref{fig:repair}.
We show the average distance between token edits (localization), average n-gram similarity between randomly sampled data points (diversity), and average token edit distance between the noisy and correct code (complexity).
Diffusion-generated data has more diversity, higher complexity, and more global errors.
The diffusion model generates both the code and the error from pure noise, whereas GPT-4o starts from provided code.

\begin{table}[htb]
\centering
\small
\caption{Results on fine-tuning different language models on the synthetic repair data generated by diffusion, GPT-4.1 and Syntactic systems (BIFI \citep{bifi} for Python and \texttt{\small flame} \citep{flame} for Excel). We see that diffusion-generated data consistently performs better than language model and syntactic systems.\\}
\label{tab:synthetic-data}
\begin{tabular}{llcccccc}
\toprule
                             &                                      &       \multicolumn{2}{c}{Python}         & \multicolumn{2}{c}{PowerShell}            & \multicolumn{2}{c}{Excel formula}  \\ \cmidrule(lr){1-2} \cmidrule(lr){3-4} \cmidrule(lr){5-6} \cmidrule(lr){7-8}
Generator     & Repair model  & Sketch & Exec. & Sketch & Exec. & Sketch & Exec. \\ \midrule
\multirow{3}{*}{Diffusion}  & CodeT5+ (2B)  & \underline{\textbf{89.2}} & \underline{\textbf{91.1}} & 25.4 & \underline{\textbf{34.2}} & 72.0 & \underline{\textbf{77.6}} \\
                            & Phi-3.5-mini-instruct (3.8B)     & 87.5 & 88.3 & \underline{\textbf{28.2}} & 33.2 & 71.0 & 76.8 \\
                            & Mistral (7B) & 87.1 & 89.3 & 27.4 & \underline{\textbf{34.2}} & \underline{\textbf{73.3}} & 75.6 \\
                            \midrule
\multirow{3}{*}{GPT-4.1} & CodeT5+ (2B)  & 87.6 & 88.2 & 23.4 & 28.1 & 69.2 & 72.1 \\
                            & Phi-3.5-mini (3.8B)    & 84.2 & 86.9 & 21.0 & 27.3 & 70.1 & 74.3 \\
                            & Mistral (7B) & 85.4 & 87.7 & 24.5 & 29.4 & 69.3 & 70.0 \\
                            \midrule
\multirow{3}{*}{Syntactic}  & CodeT5+ (2B)  & 85.4 & 87.3 & --  & --  & 70.1 & 72.4 \\
                            & Phi-3.5-mini (3.8B)    & 84.2 & 85.3 & --  & --  & 72.4 & \underline{\textbf{77.6}} \\
                            & Mistral (7B) & 86.0 & 89.3 & --  & --  & 71.2 & 73.5 \\
                            \bottomrule
\end{tabular}
\end{table}

\begin{figure}[tbh]
    \centering
    \begin{subfigure}[b]{0.3\textwidth}
        \centering
        \includegraphics[width=\textwidth]{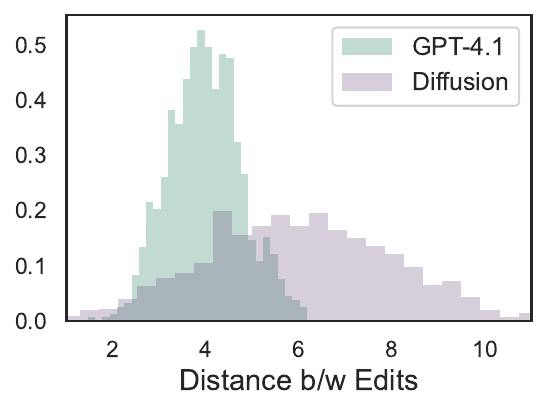}
        \caption{Localization}
        \label{fig:syntehtic-complexity}
    \end{subfigure}
    \hfill 
    \begin{subfigure}[b]{0.3\textwidth}
        \centering
        \includegraphics[width=\textwidth]{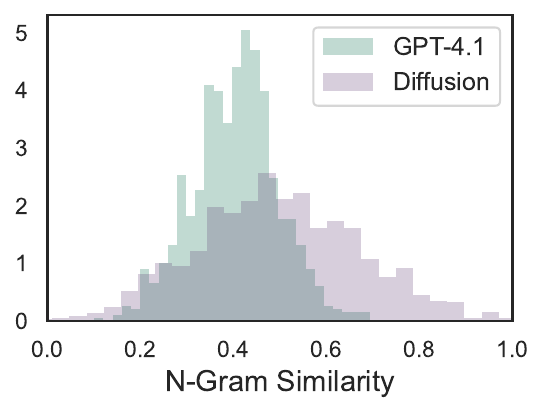}
        \caption{Diversity}
        \label{fig:synthetic-diversity}
    \end{subfigure}
    \hfill 
    \begin{subfigure}[b]{0.3\textwidth}
        \centering
        \includegraphics[width=\textwidth]{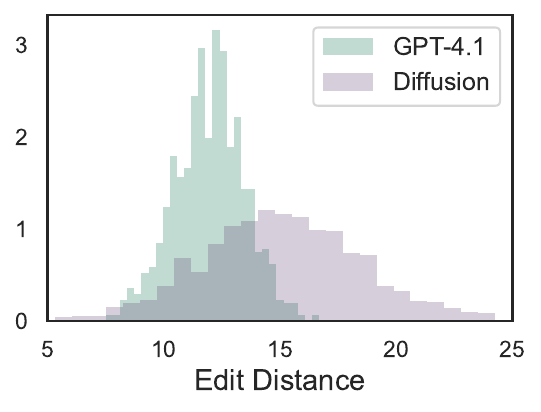}
        \caption{Complexity}
        \label{fig:synthetic-edit}
    \end{subfigure}
    \caption{Figure showing the trends in the diffusion (purple) and GPT-4.1 (green) generated repair data. We show (a) Localization---average distance between edits; (b) Diversity---average n-gram diversity in generated correct code; and (c) Complexity---edit distance of the repair.}
    \label{fig:synthetic-analysis}
\end{figure}

\begin{figure}
    \centering
    \includegraphics[width=0.65\linewidth]{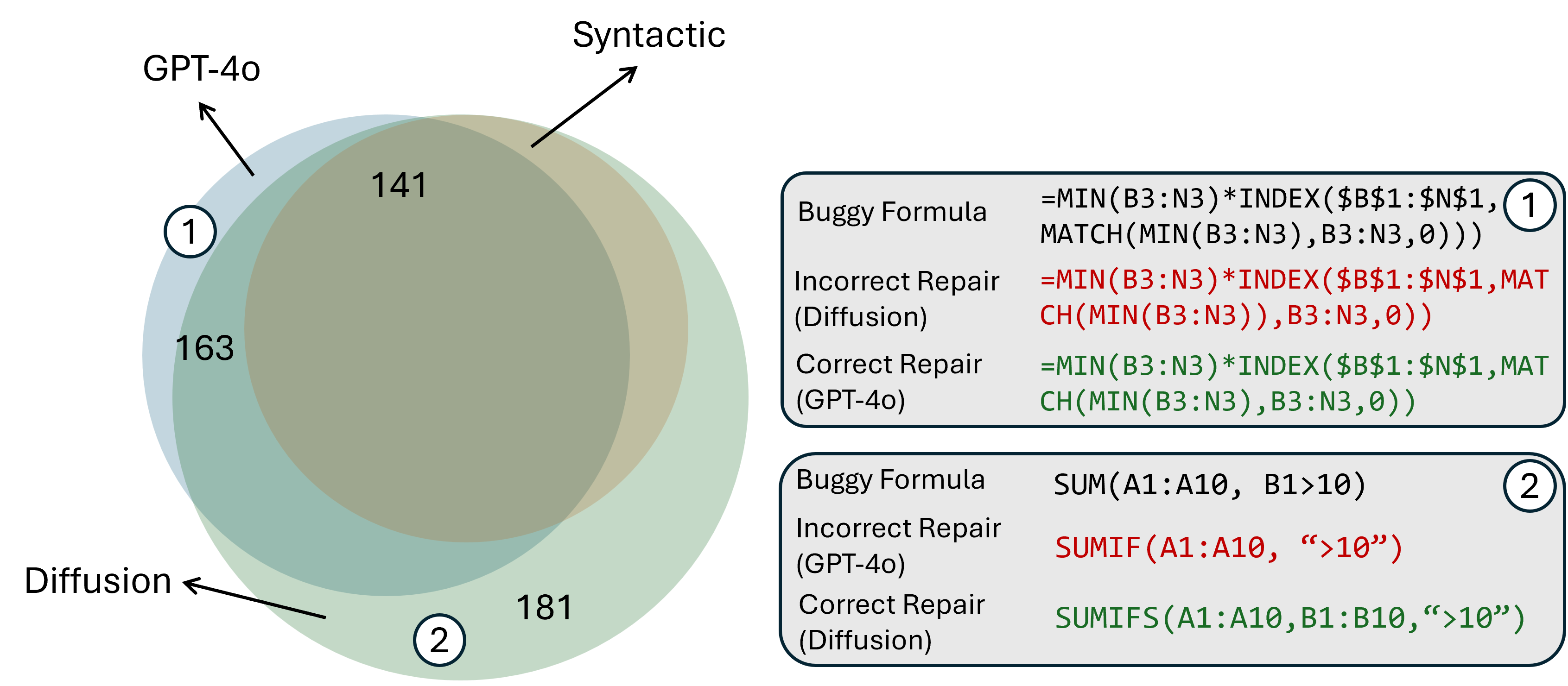}
    \caption{Venn diagram of benchmarks solved correctly for models trained on synthetic datasets generated from different sources.
    The example shows cases where diffusion data trained model is able to repair a task while GPT-4o data trained model cannot and vice versa.
    }
    \label{fig:venn}
\end{figure}

Finally, Figure~\ref{fig:venn} shows the overlap between Excel benchmarks solved using the CodeT5+ model for different sources of synthetic data.
Using diffusion data solves all cases that are solved by the synthetic data, which required manual analysis of human errors to manually implement 17 noise operators.
Bigger mistakes, like completely missing an argument spanning multiple tokens, occur more in the diffusion data.
An extra parenthesis does not occur as much in the diffusion data, as the pre-trained models quickly learns this structure, and is an explicit instruction in the GPT-4o prompt.

\section{Related work}

\textbf{Diffusion models for text and code}
Diffusion models have shown their ability to gradually refine noisy data into realistic outputs through a denoising process \citep{diffusionoriginal}.
They were originally popularized to generate photo-realistic images \citep{diffusion,diffusionhigh} and later applied to other high-dimensional data generation, like audio \citep{diffusionaudio} and video \citep{videodiffusion} synthesis.
Diffusion has also been adapted for to discrete domains like text \citep{diffusionlm, genie} and code \citep{codefusion} where the ability to look at the whole previous generation has benefits over auto-regressive generation.
Two approaches are embedding discrete tokens into a continuous space where the diffusion takes place and then decoding \cite{diffusionlm} or directly performing diffusion in the discrete space through a transition matrix \cite{diffusionbert}.

\textbf{Code repair}
Automated code repair \citep{repairsurvey} has long been a key challenge in software engineering, with early approaches using heuristic searches \citep{aprrandomsearch} and program synthesis \citep{semfix,lamirage}.
More recently, transformer-based systems have been shown adept at learning to repair code \citep{tfix,bifi,aprnmt,khatry2023wordscodeharnessingdata, vehicle-tele}.
A major limitation of training a repair model is the requirement for large quantities of data.
That is not true anymore for large language models, which are adept at repair code and can take in additional context like error messages \citep{ring}.
In order to leverage the corpora of unsupervised data, previous works have explored generating synthetic data using static rules \citep{flame,deepfix,great} or learning to break programs in a \emph{natural} way \citep{aprselfsupervised,bifi}.
These approaches are limited to the encoded rules and the quality of the learned code breaker, and thus suffer from out-of-domain generalization.


\section{Conclusion and limitations}

In this paper, we explored the potential of applying pre-trained code diffusion to the problem of last-mile repair.
These diffusion models iteratively denoise a latent representation of code and the discrete decoding of intermediate steps resemble last-mile programming errors.
Our experiments show that injecting actual broken code into this process can cause the diffusion process to repair the code, and that sampling these intermediate step yields data that can be used to fine-tune last-mile repair models.
In its current state, using diffusion models to generate synthetic training data shows the most promise.

Diffusion for code has only been applied to shorter snippets with smaller models, on relatively small datasets.
Since there is no additional context, like error messages or test cases, the model might not capture some of the semantics of the broken snippet.
Our findings consider the diffusion model as-is: controlled decoding \citep{diffusionlm, format5} can help in remaining close to the source snippet.

\bibliography{v2/references}

\begin{thebibliography}{35}
\providecommand{\natexlab}[1]{#1}
\providecommand{\url}[1]{\texttt{#1}}
\expandafter\ifx\csname urlstyle\endcsname\relax
  \providecommand{\doi}[1]{doi: #1}\else
  \providecommand{\doi}{doi: \begingroup \urlstyle{rm}\Url}\fi

\bibitem[Ho et~al.(2020)Ho, Jain, and Abbeel]{diffusion}
Jonathan Ho, Ajay Jain, and Pieter Abbeel.
\newblock Denoising diffusion probabilistic models.
\newblock \emph{Advances in neural information processing systems}, 33:\penalty0 6840--6851, 2020.

\bibitem[Ho et~al.(2022)Ho, Salimans, Gritsenko, Chan, Norouzi, and Fleet]{videodiffusion}
Jonathan Ho, Tim Salimans, Alexey Gritsenko, William Chan, Mohammad Norouzi, and David~J Fleet.
\newblock Video diffusion models.
\newblock \emph{Advances in Neural Information Processing Systems}, 35:\penalty0 8633--8646, 2022.

\bibitem[Xing et~al.(2023)Xing, Feng, Chen, Dai, Hu, Xu, Wu, and Jiang]{videodiffusionsurvey}
Zhen Xing, Qijun Feng, Haoran Chen, Qi~Dai, Han Hu, Hang Xu, Zuxuan Wu, and Yu-Gang Jiang.
\newblock A survey on video diffusion models.
\newblock \emph{ACM Computing Surveys}, 2023.

\bibitem[Li et~al.(2022)Li, Thickstun, Gulrajani, Liang, and Hashimoto]{diffusionlm}
Xiang Li, John Thickstun, Ishaan Gulrajani, Percy~S Liang, and Tatsunori~B Hashimoto.
\newblock Diffusion-lm improves controllable text generation.
\newblock \emph{Advances in Neural Information Processing Systems}, 35:\penalty0 4328--4343, 2022.

\bibitem[Singh et~al.(2023{\natexlab{a}})Singh, Cambronero, Gulwani, Le, Negreanu, and Verbruggen]{codefusion}
Mukul Singh, Jos{\'e} Cambronero, Sumit Gulwani, Vu~Le, Carina Negreanu, and Gust Verbruggen.
\newblock Codefusion: A pre-trained diffusion model for code generation.
\newblock In \emph{Proceedings of the 2023 Conference on Empirical Methods in Natural Language Processing}, pages 11697--11708, 2023{\natexlab{a}}.

\bibitem[Lin et~al.(2023)Lin, Gong, Shen, Wu, Fan, Lin, Duan, and Chen]{genie}
Zhenghao Lin, Yeyun Gong, Yelong Shen, Tong Wu, Zhihao Fan, Chen Lin, Nan Duan, and Weizhu Chen.
\newblock Text generation with diffusion language models: A pre-training approach with continuous paragraph denoise.
\newblock In \emph{International Conference on Machine Learning}, pages 21051--21064. PMLR, 2023.

\bibitem[Bavishi et~al.(2022)Bavishi, Joshi, Cambronero, Fariha, Gulwani, Le, Radi{\v{c}}ek, and Tiwari]{lamirage}
Rohan Bavishi, Harshit Joshi, Jos{\'e} Cambronero, Anna Fariha, Sumit Gulwani, Vu~Le, Ivan Radi{\v{c}}ek, and Ashish Tiwari.
\newblock Neurosymbolic repair for low-code formula languages.
\newblock \emph{Proceedings of the ACM on Programming Languages}, 6\penalty0 (OOPSLA2):\penalty0 1093--1122, 2022.

\bibitem[Huang et~al.(2023)Huang, Meng, Zhang, Liu, Wang, Li, and Zhang]{repairfinetuning}
Kai Huang, Xiangxin Meng, Jian Zhang, Yang Liu, Wenjie Wang, Shuhao Li, and Yuqing Zhang.
\newblock An empirical study on fine-tuning large language models of code for automated program repair.
\newblock In \emph{2023 38th IEEE/ACM International Conference on Automated Software Engineering (ASE)}, pages 1162--1174. IEEE, 2023.

\bibitem[Joshi et~al.(2024)Joshi, Ebenezer, Sanchez, Gulwani, Kanade, Le, Radi{\v{c}}ek, and Verbruggen]{flame}
Harshit Joshi, Abishai Ebenezer, Jos{\'e}~Cambronero Sanchez, Sumit Gulwani, Aditya Kanade, Vu~Le, Ivan Radi{\v{c}}ek, and Gust Verbruggen.
\newblock Flame: A small language model for spreadsheet formulas.
\newblock In \emph{Proceedings of the AAAI Conference on Artificial Intelligence}, volume~38, pages 12995--13003, 2024.

\bibitem[Joshi et~al.(2023)Joshi, Sanchez, Gulwani, Le, Verbruggen, and Radi{\v{c}}ek]{ring}
Harshit Joshi, Jos{\'e}~Cambronero Sanchez, Sumit Gulwani, Vu~Le, Gust Verbruggen, and Ivan Radi{\v{c}}ek.
\newblock Repair is nearly generation: Multilingual program repair with llms.
\newblock In \emph{Proceedings of the AAAI Conference on Artificial Intelligence}, volume~37, pages 5131--5140, 2023.

\bibitem[Yasunaga and Liang(2020)]{aprselfsupervised}
Michihiro Yasunaga and Percy Liang.
\newblock Graph-based, self-supervised program repair from diagnostic feedback.
\newblock In \emph{International Conference on Machine Learning}, pages 10799--10808. PMLR, 2020.

\bibitem[Singh et~al.(2025)Singh, Cambronero, Gulwani, Le, Negreanu, Radhakrishna, and Verbruggen]{datavinci}
Mukul Singh, Jos\'{e} Cambronero, Sumit Gulwani, Vu~Le, Carina Negreanu, Arjun Radhakrishna, and Gust Verbruggen.
\newblock Datavinci: Learning syntactic and semantic string repairs.
\newblock \emph{Proc. ACM Manag. Data}, 3\penalty0 (1), February 2025.
\newblock \doi{10.1145/3709677}.
\newblock URL \url{https://doi.org/10.1145/3709677}.

\bibitem[Singha et~al.(2024)Singha, Chopra, Khatry, Gulwani, Henley, Le, Parnin, Singh, and Verbruggen]{aligned-code-gen}
Ananya Singha, Bhavya Chopra, Anirudh Khatry, Sumit Gulwani, Austin Henley, Vu~Le, Chris Parnin, Mukul Singh, and Gust Verbruggen.
\newblock Semantically aligned question and code generation for automated insight generation.
\newblock In \emph{Proceedings of the 1st International Workshop on Large Language Models for Code}, LLM4Code '24, page 127–134, New York, NY, USA, 2024. Association for Computing Machinery.
\newblock ISBN 9798400705793.
\newblock \doi{10.1145/3643795.3648381}.
\newblock URL \url{https://doi.org/10.1145/3643795.3648381}.

\bibitem[Yasunaga and Liang(2021)]{bifi}
Michihiro Yasunaga and Percy Liang.
\newblock Break-it-fix-it: Unsupervised learning for program repair.
\newblock In \emph{International conference on machine learning}, pages 11941--11952. PMLR, 2021.

\bibitem[Singh et~al.(2022{\natexlab{a}})Singh, Cambronero, Gulwani, Le, Negreanu, Raza, and Verbruggen]{cornet}
Mukul Singh, Jos{\'e} Cambronero, Sumit Gulwani, Vu~Le, Carina Negreanu, Mohammad Raza, and Gust Verbruggen.
\newblock Cornet: A neurosymbolic approach to learning conditional table formatting rules by example.
\newblock \emph{arXiv preprint arXiv:2208.06032}, 2022{\natexlab{a}}.

\bibitem[Payan et~al.(2023)Payan, Mishra, Singh, Negreanu, Poelitz, Baral, Roy, Chakravarthy, Durme, and Nouri]{payan2023instructexcelbenchmarknaturallanguage}
Justin Payan, Swaroop Mishra, Mukul Singh, Carina Negreanu, Christian Poelitz, Chitta Baral, Subhro Roy, Rasika Chakravarthy, Benjamin~Van Durme, and Elnaz Nouri.
\newblock Instructexcel: A benchmark for natural language instruction in excel, 2023.
\newblock URL \url{https://arxiv.org/abs/2310.14495}.

\bibitem[Vaswani et~al.(2017)Vaswani, Shazeer, Parmar, Uszkoreit, Jones, Gomez, Kaiser, and Polosukhin]{transformer}
Ashish Vaswani, Noam Shazeer, Niki Parmar, Jakob Uszkoreit, Llion Jones, Aidan~N Gomez, \L~ukasz Kaiser, and Illia Polosukhin.
\newblock Attention is all you need.
\newblock In I.~Guyon, U.~Von Luxburg, S.~Bengio, H.~Wallach, R.~Fergus, S.~Vishwanathan, and R.~Garnett, editors, \emph{Advances in Neural Information Processing Systems}, volume~30. Curran Associates, Inc., 2017.
\newblock URL \url{https://proceedings.neurips.cc/paper/2017/file/3f5ee243547dee91fbd053c1c4a845aa-Paper.pdf}.

\bibitem[Chen et~al.(2021)Chen, Tworek, Jun, Yuan, Pinto, Kaplan, Edwards, Burda, Joseph, Brockman, et~al.]{codex}
Mark Chen, Jerry Tworek, Heewoo Jun, Qiming Yuan, Henrique Ponde de~Oliveira Pinto, Jared Kaplan, Harri Edwards, Yuri Burda, Nicholas Joseph, Greg Brockman, et~al.
\newblock Evaluating large language models trained on code.
\newblock \emph{arXiv preprint arXiv:2107.03374}, 2021.

\bibitem[Wang et~al.(2023)Wang, Le, Gotmare, Bui, Li, and Hoi]{codet5+}
Yue Wang, Hung Le, Akhilesh~Deepak Gotmare, Nghi D.~Q. Bui, Junnan Li, and Steven C.~H. Hoi.
\newblock Codet5+: Open code large language models for code understanding and generation, 2023.

\bibitem[Abdin et~al.(2024)Abdin, Aneja, Awadalla, Awadallah, Awan, Bach, Bahree, Bakhtiari, Bao, Behl, et~al.]{phi3}
Marah Abdin, Jyoti Aneja, Hany Awadalla, Ahmed Awadallah, Ammar~Ahmad Awan, Nguyen Bach, Amit Bahree, Arash Bakhtiari, Jianmin Bao, Harkirat Behl, et~al.
\newblock Phi-3 technical report: A highly capable language model locally on your phone.
\newblock \emph{arXiv preprint arXiv:2404.14219}, 2024.

\bibitem[Jiang et~al.(2023)Jiang, Sablayrolles, Mensch, Bamford, Chaplot, Casas, Bressand, Lengyel, Lample, Saulnier, et~al.]{mistral}
Albert~Q Jiang, Alexandre Sablayrolles, Arthur Mensch, Chris Bamford, Devendra~Singh Chaplot, Diego de~las Casas, Florian Bressand, Gianna Lengyel, Guillaume Lample, Lucile Saulnier, et~al.
\newblock Mistral 7b.
\newblock \emph{arXiv preprint arXiv:2310.06825}, 2023.

\bibitem[Sohl-Dickstein et~al.(2015)Sohl-Dickstein, Weiss, Maheswaranathan, and Ganguli]{diffusionoriginal}
Jascha Sohl-Dickstein, Eric Weiss, Niru Maheswaranathan, and Surya Ganguli.
\newblock Deep unsupervised learning using nonequilibrium thermodynamics.
\newblock In \emph{International conference on machine learning}, pages 2256--2265. PMLR, 2015.

\bibitem[Rombach et~al.(2022)Rombach, Blattmann, Lorenz, Esser, and Ommer]{diffusionhigh}
Robin Rombach, Andreas Blattmann, Dominik Lorenz, Patrick Esser, and Bj{\"o}rn Ommer.
\newblock High-resolution image synthesis with latent diffusion models.
\newblock In \emph{Proceedings of the IEEE/CVF conference on computer vision and pattern recognition}, pages 10684--10695, 2022.

\bibitem[Kong et~al.(2021)Kong, Ping, Huang, Zhao, and Catanzaro]{diffusionaudio}
Zhifeng Kong, Wei Ping, Jiaji Huang, Kexin Zhao, and Bryan Catanzaro.
\newblock Diffwave: A versatile diffusion model for audio synthesis.
\newblock In \emph{International Conference on Learning Representations}, 2021.

\bibitem[He et~al.(2023)He, Sun, Tang, Wang, Huang, and Qiu]{diffusionbert}
Zhengfu He, Tianxiang Sun, Qiong Tang, Kuanning Wang, Xuan-Jing Huang, and Xipeng Qiu.
\newblock Diffusionbert: Improving generative masked language models with diffusion models.
\newblock In \emph{Proceedings of the 61st Annual Meeting of the Association for Computational Linguistics}, pages 4521--4534, 2023.

\bibitem[Zhang et~al.(2023)Zhang, Fang, Ma, Sun, and Chen]{repairsurvey}
Quanjun Zhang, Chunrong Fang, Yuxiang Ma, Weisong Sun, and Zhenyu Chen.
\newblock A survey of learning-based automated program repair.
\newblock \emph{ACM Transactions on Software Engineering and Methodology}, 33\penalty0 (2):\penalty0 1--69, 2023.

\bibitem[Qi et~al.(2014)Qi, Mao, Lei, Dai, and Wang]{aprrandomsearch}
Yuhua Qi, Xiaoguang Mao, Yan Lei, Ziying Dai, and Chengsong Wang.
\newblock The strength of random search on automated program repair.
\newblock In \emph{Proceedings of the 36th International Conference on Software Engineering}, pages 254--265, 2014.

\bibitem[Nguyen et~al.(2013)Nguyen, Qi, Roychoudhury, and Chandra]{semfix}
Hoang D.~T. Nguyen, Dawei Qi, Abhik Roychoudhury, and Satish Chandra.
\newblock Semfix: Program repair via semantic analysis.
\newblock \emph{International Conference on Software Engineering}, pages 772--781, 2013.

\bibitem[Berabi et~al.(2021)Berabi, He, Raychev, and Vechev]{tfix}
Berkay Berabi, Jingxuan He, Veselin Raychev, and Martin Vechev.
\newblock Tfix: Learning to fix coding errors with a text-to-text transformer.
\newblock In \emph{International Conference on Machine Learning}, pages 780--791. PMLR, 2021.

\bibitem[Tufano et~al.(2019)Tufano, Watson, Bavota, Penta, White, and Poshyvanyk]{aprnmt}
Michele Tufano, Cody Watson, Gabriele Bavota, Massimiliano~Di Penta, Martin White, and Denys Poshyvanyk.
\newblock An empirical study on learning bug-fixing patches in the wild via neural machine translation.
\newblock \emph{ACM Transactions on Software Engineering and Methodology (TOSEM)}, 28\penalty0 (4):\penalty0 1--29, 2019.

\bibitem[Khatry et~al.(2023)Khatry, Cahoon, Henkel, Deep, Emani, Floratou, Gulwani, Le, Raza, Shi, Singh, and Tiwari]{khatry2023wordscodeharnessingdata}
Anirudh Khatry, Joyce Cahoon, Jordan Henkel, Shaleen Deep, Venkatesh Emani, Avrilia Floratou, Sumit Gulwani, Vu~Le, Mohammad Raza, Sherry Shi, Mukul Singh, and Ashish Tiwari.
\newblock From words to code: Harnessing data for program synthesis from natural language, 2023.
\newblock URL \url{https://arxiv.org/abs/2305.01598}.

\bibitem[Singh et~al.(2022{\natexlab{b}})Singh, Dubey, and Kumar]{vehicle-tele}
Mukul Singh, Rahul~Kumar Dubey, and Swarup Kumar.
\newblock Chapter 15 - vehicle telematics: An internet of things and big data approach.
\newblock In Rajiv Pandey, Sunil~Kumar Khatri, Neeraj kumar Singh, and Parul Verma, editors, \emph{Artificial Intelligence and Machine Learning for EDGE Computing}, pages 235--254. Academic Press, 2022{\natexlab{b}}.
\newblock ISBN 978-0-12-824054-0.
\newblock \doi{https://doi.org/10.1016/B978-0-12-824054-0.00019-8}.
\newblock URL \url{https://www.sciencedirect.com/science/article/pii/B9780128240540000198}.

\bibitem[Gupta et~al.(2017)Gupta, Pal, Kanade, and Shevade]{deepfix}
Rahul Gupta, Soham Pal, Aditya Kanade, and Shirish Shevade.
\newblock Deepfix: Fixing common c language errors by deep learning.
\newblock In \emph{Proceedings of the AAAI Conference on Artificial Intelligence}, volume~31, 2017.

\bibitem[Hellendoorn et~al.(2019)Hellendoorn, Sutton, Singh, Maniatis, and Bieber]{great}
Vincent~J Hellendoorn, Charles Sutton, Rishabh Singh, Petros Maniatis, and David Bieber.
\newblock Global relational models of source code.
\newblock In \emph{International conference on learning representations}, 2019.

\bibitem[Singh et~al.(2023{\natexlab{b}})Singh, Cambronero, Gulwani, Le, Negreanu, Nouri, Raza, and Verbruggen]{format5}
Mukul Singh, Jos\'{e} Cambronero, Sumit Gulwani, Vu~Le, Carina Negreanu, Elnaz Nouri, Mohammad Raza, and Gust Verbruggen.
\newblock Format5: Abstention and examples for conditional table formatting with natural language.
\newblock 17\penalty0 (3):\penalty0 497–510, November 2023{\natexlab{b}}.
\newblock ISSN 2150-8097.
\newblock \doi{10.14778/3632093.3632111}.
\newblock URL \url{https://doi.org/10.14778/3632093.3632111}.

\end{thebibliography}
\bibliographystyle{unsrtnat}

\end{document}